\documentclass[runningheads]{svmult}

\usepackage{makeidx}   
\usepackage{graphicx}  
\usepackage{subeqnar}  
\usepackage{multicol}  
\usepackage{physmubb}  
\makeindex             

\usepackage{wasysym}   
\begin{document}
\title*{Frequency Dependent Electrical Transport \\ 
in the Integer Quantum Hall Effect}
\toctitle{Frequency dependent electrical transport in 
\protect\newline the integer quantum Hall effect}
%
%
\titlerunning{Frequency dependent QHE}
%
\author{Ludwig Schweitzer}

%
%
%

\maketitle              

\section{Introduction}
It is well established to view the integer quantum Hall effect (QHE) as 
a sequence of quantum phase transitions associated with critical points   
that separate energy regions of localised states where the Hall-conductivity 
$\sigma_{xy}$ is quantised in integer units of $e^2/h$ (see, e.g., 
\cite{Huc95,SGCS97}). Simultaneously, the longitudinal conductivity 
$\sigma_{xx}$ gets unmeasurable small in the limit of vanishing 
temperature and zero frequency. To check the inherent consequences of 
this theoretical picture, various experiments have been devised to
investigate those properties that should occur near the critical 
energies $E_n$ assigned to the critical points. For example, due to 
the divergence 
of the localisation length $\xi(E)\propto |E-E_n|^{-\mu}$, the width 
$\Delta$ of the longitudinal conductivity peaks emerging at the  
transitions is expected to exhibit power-law scaling with respect 
to temperature, system size, or an externally applied frequency. 

High frequency Hall-conductivity experiments, initially aimed
at resolving the problem of the so-called low frequency breakdown of
the QHE apparently observed at $\sim 1$\,MHz, were successfully
carried out at microwave frequencies ($\sim 33$\,MHz) \cite{KMWS86}.
The longitudinal ac conductivity was also studied to obtain some 
information about localisation and the formation of Hall plateaus
in the frequency ranges 100\,Hz to 20\,kHz \cite{LGHS87} and 
50--600\,MHz \cite{BPRT90}.   
Later, frequency dependent transport has been investigated also in the 
Gigahertz frequency range below 15\,GHz \cite{ESKT93,BMB98} and above 
30\,GHz \cite{BMKHP98,MKBHP98,KMK99}. 

Dynamical scaling has been studied in several experiments, 
some of which show indeed power law scaling of the
$\sigma_{xx}(\omega)$ peak width as expected, 
$\Delta\sim\omega^{\kappa}$ \cite{ESKT93,SET94,HZHMDK02}, 
whereas others do not \cite{BMB98}. The exponent $\kappa=(\mu z)^{-1}$
contains both the critical exponent $\mu$ of the localisation length 
and the dynamical exponent $z$ which relates energy and length scales,
$E\sim L^{-z}$. The value of $\mu=2.35\pm 0.03$ is well known from
numerical calculations \cite{HK90,Huc92}, and it also coincides with 
the outcome of a finite size scaling experiment \cite{KHKP91}. However, 
it is presently only accepted as true that $z=2$ for non-interacting 
particles, and $z=1$ if Coulomb electron-electron interactions are 
present \cite{LW96,YM93,HB99,WFGC00,WX02}.
Therefore, a theoretical description of the ac conductivities would 
clearly contribute to a better understanding of dynamical scaling at 
quantum critical points.

Legal metrology represents a second area where a better knowledge 
of frequency dependent transport is highly desirable because 
the ac quantum Hall effect is applied for the realization and
dissemination of the impedance standard and the unit of capacitance, 
the farad. At the moment the achieved relative uncertainty at a 
frequency of 1\,kHz is of the order $10^{-7}$ which still is at 
least one order of magnitude to large \cite{Del93,Del94,CHK99,SMCHAP02}. 
It is unclear whether the observed deviations from the quantised dc value 
are due to external influences like capacitive and inductive couplings 
caused by the leads and contacts. Alternatively, the measured
frequency effects that make exact quantisation impossible
could be already inherent in an ideal non-interacting two-dimensional 
electron gas in the presence of disorder and a perpendicular magnetic field. 
Of course, applying a finite frequency $\omega$ will lead to a finite
$\sigma_{xx}(\omega)$ and in turn will influence the quantisation of 
$\sigma_{xy}(\omega)$, but it remains to be investigated how large the
deviation will be.  

Theoretical studies of the ac conductivity in quantum Hall systems started
in 1985 \cite{Joy85} when it was shown within a semiclassical percolation 
theory that for finite frequencies the longitudinal conductivity is not  
zero, thus influencing the quantisation of the Hall-conductivity.
The quantum mechanical problem of non-interacting electrons in a 2d
disordered system in the presence of a strong perpendicular magnetic field
$B$ was tackled by Apel \cite{Ape89} using a variational method. Applying an 
instanton approximation and confining to the high field limit, i.e., restricting 
to the lowest Landau level (LLL), an analytical solution for the real part of
the frequency dependent longitudinal conductivity could be presented, 
$\sigma_{xx}(\omega)\propto\omega^2\ln(1/\omega^2)$,
a result that should hold if the Fermi energy lies deep down in the lower
tail of the LLL.

Generalising the above result, both real and imaginary parts of the
frequency dependent conductivities were obtained in a sequence of
papers by Viehweger and Efetov \cite{VE90,VE91,VE91a}. The Kubo 
conductivities were determined by calculating the functional integrals
in super-symmetric representation near non-trivial saddle points.
Still, for the final results the Fermi energy was restricted 
to lie within the energy range of localised states in the lowest 
tail of the lowest Landau band and, therefore, no proposition for 
the critical regions at half fillings could be given. The longitudinal
conductivity was found to be  
\begin{equation}
\sigma_{xx}(\omega)=
c\,\omega^2 \ln^b(1/\omega^2)-\I e^2\omega 2l_{B}^2\varrho(E_{\mathrm F})\;,
\label{VEsxx}
\end{equation}
with the density of states $\varrho(E_{\mathrm F})$, and two unspecified
constants $b$ and $c$ \cite{VE90}.
The real part of the Hall-conductivity in the same limit was proposed as
\begin{equation}
\sigma_{yx}(\omega)=e^2/h\,(\hbar\omega/\Gamma)^2\,8\pi l^2_{B}n_e\;,
\end{equation}
where $l_{B}=\sqrt{\hbar/(eB)}$ is the magnetic length,
$\Gamma^2=\lambda/2\pi l^2_{B}$ is the second moment of the white 
noise disorder potential distribution with disorder strength $\lambda$, 
and $n_e$ denotes the electron density. 
The deviation due to frequency of the Hall-conductivity from its 
quantised dc plateau value can be perceived from an approximate expression
proposed by Viehweger and Efetov \cite{VE91a} for the 2nd plateau, 
e.g., for filling factor $\nu\approx 2$,  
\begin{equation}
\sigma_{yx}(\omega)=e^2/h\,\Big[\frac{2}{1-\omega^2/\omega_c^2}-       
\Big(\frac{2\hbar\omega}{\Gamma}\Big)^2\,(2-\nu)\Big]\;.
\label{delsxy}
\end{equation}
Again, $\Gamma$ is a measure of the disorder strength describing
the width of the disorder broadened Landau band, and $\omega_c=eB/m_e$
is the cyclotron frequency with electron mass $m_e$. According to 
(\ref{delsxy}), due to frequency a deviation from the quantised 
value becomes apparent even for integer filling. 

Before reviewing the attempts which applied numerical methods to       
overcome the limitations that had to be conceded in connexion with 
the position of $E_{\mathrm F}$ in the analytical work, and to check    
the permissiveness of the approximations made, it is appropriate here 
to mention a result for the hopping regime.
Polyakov and Shklovskii \cite{PS93a} obtained for the dissipative part
of the ac conductivity a relation which, in contrast to (\ref{VEsxx}), 
is linear in frequency $\omega$,
\begin{equation}
\Re\,\sigma_{xx}(\omega)=K\,\epsilon\xi\omega\;.
\end{equation} 
This expression has recently been used to successfully describe 
experimental data \cite{HZH01}. Here, $\xi$ is the localisation 
length, $\epsilon$ the dielectric constant, and the pre-factor is 
$K=1/6$ in the limit $\hbar\omega\gg k_{\mathrm B}T$. Also, the
frequency scaling of the peak width 
$\Delta \sim \omega^{\kappa}$
was proposed within the same hopping model \cite{PS93a}. 

Turning now to the numerical approaches which were started by
Gammel and Brenig who considered the low frequency anomalies 
and the finite size scaling of the real part of the conductivity peak 
$\sigma^p_{xx}(\omega,L_y)$ at the critical point of the lowest Landau
band \cite{GB96}. For these purposes the authors utilised the random Landau
model in the high field limit (lowest Landau band only) \cite{Huc95,HK90}  
and generalised MacKinnon's recursive Green function method \cite{Mac85} 
for the evaluation of the real part of the dynamical conductivity. 
In contrast to the conventional quadratic Drude-like behaviour the
peak value decayed linearly with frequency, 
$\sigma^p_{xx}(\omega)=\sigma^0_{xx}-\mathrm{const.}\,|\omega|$, 
which was attributed to the long time tails in the velocity correlations 
which were observed also in a semiclassical model \cite{EB94,BGK97}. 
The range of this unusual linear frequency dependence varied with the 
spatial correlation length of the disorder potentials. A second result
concerns the scaling of $\sigma^p_{xx}(\omega,L_y)$ at low frequencies
as a function of the system width $L_{y}$, 
$\sigma^p_{xx}(\omega L^2_y)\propto (\omega L^2_y)^{2-\eta/2}$, where 
$\eta=2-D(2)=0.36\pm 0.06$ \cite{GB96} is related to the multi-fractal
wave functions \cite{HS92,PJ91,HKS92,Jan94}, and to the anomalous
diffusion at the critical point with $\eta\simeq 0.38$ \cite{CD88} and 
$D(2)\simeq 1.62$ \cite{HS94}.

The frequency scaling of the $\sigma_{xx}(\omega)$ peak width was considered 
numerically for the first time in a paper by Avishai and Luck \cite{AL96}. 
Using a continuum model with spatially correlated Gaussian disorder
potentials placed on a square lattice, which then was diagonalised within 
the subspace of functions pertaining to the lowest Landau level, the
real part of the dissipative conductivity was evaluated from the Kubo
formula involving matrix elements of the velocity of the guiding centres
\cite{And84a}. This is always necessary in single band approximations 
because of the vanishing of the current matrix elements between states 
belonging to the same Landau level. As a result, a broadening of the 
conductivity peak was observed and from a finite size scaling analysis
a dynamical exponent $z=1.19\pm0.13$ could be extracted using
$\nu=2.33$ from \cite{HK90}. This is rather startling because it 
is firmly believed that for non-interacting systems $z$ equals the 
Euclidean dimension of space which gives $z=d=2$ in the QHE case.

A different theoretical approach for the low frequency behaviour of the
$\sigma_{xx}(\omega)$ peak has been pursued by Jug and Ziegler \cite{JZ99} 
who studied a Dirac fermion model with an inhomogeneous mass \cite{Lea94} 
applying a non-perturbative calculation. This model leads to a non-zero 
density of states and to a finite bandwidth of extended states near 
the centre of the Landau band \cite{Zie97}.
Therefore, the ac conductivity as well as its peak width do neither show 
power-law behaviour nor do they vanish in the limit $\omega\to 0$. 
This latter feature of the model has been asserted to explain the linear
frequency dependence and the finite intercept at $\omega=0$ observed 
experimentally for the width of the conductivity peak by Balaban 
\textit{et al.} \cite{BMB98}, but, up to now, there is no other
experiment showing such a peculiar behaviour.

\section{Preliminary considerations -- basic relations}
In the usual experiments on two-dimensional systems a current 
$I_{x}(\omega)$ is driven through the sample of length $L_{x}$ and
width $L_{y}$. The voltage drop along the current direction, 
$U_{x}(\omega)$, and that across the sample, $U_{y}(\omega)$,
are measured from which the Hall-resistance 
$R_{\mathrm H}(\omega)=U_{y}(\omega)/I_{x}(\omega)=\rho_{xy}(\omega)$  
and the longitudinal resistance
$R_{x}(\omega)=U_{x}(\omega)/I_{x}(\omega)=\rho_{xx}(\omega)L_{x}/L_{y}$  
are obtained, where $\rho_{xy}$ and $\rho_{xx}$ denote the respective
resistivities. To compare with the theoretically calculated 
conductivities one has to use the relations in the following, only in Corbino
samples $\sigma_{xx}(\omega)$ can be experimentally detected directly. 

The total current through a cross-section,   
$I_{x}(\omega)=\int_{0}^{L_{y}} j_{x}(\omega,\vec{r})\D y$,  
is determined by the local current density $j_{x}(\omega,\vec{r})$ 
which constitutes the response to the applied electric field
\begin{equation}
j_{x}(\omega,\vec{r})=\int\sum_{u\in \{x,y\}}
\sigma_{xu}(\omega,\vec{r},\vec{r}') E_{u}(\omega,\vec{r}')\, \D^2\vec{r}'\;.
\end{equation}
The nonlocal conductivity tensor is particularly important in phase-coherent
mesoscopic samples. Usually, for the investigation of the measured macroscopic
conductivity tensor one is not interested in its spatial dependence. Therefore, 
one relies on a local approximation and considers the electric field
to be effectively constant. This leads to Ohm's law 
$\vec{j}=\widehat{\sigma}\vec{E}$ from which the resistance components
are simply given by inverting the conductivity tensor $\widehat{\sigma}$
\begin{equation}
\left ( \begin{array}{ll}\rho_{xx} & \rho_{xy}\\ \rho_{yx} & \rho_{yy}
\end{array}\right ) = 
\frac{1}{\sigma_{xx}\sigma_{yy}-\sigma_{xy}\sigma_{yx}}\
\left ( \begin{array}{ll}\sigma_{yy} & \sigma_{yx}\\ \sigma_{xy} & \sigma_{xx}
\end{array}\right )\;.
\end{equation}
For an isotropic system we have $\sigma_{xx}=\sigma_{yy}$ and 
$\sigma_{yx} = - \sigma_{xy}$ which in case of zero frequency gives
the well known relations
\begin{equation}
\rho_{xx}=\frac{\sigma_{xx}}{\sigma_{xx}^2+\sigma_{xy}^2},\hspace{.5cm}
\rho_{xy}=\frac{-\sigma_{xy}}{\sigma_{xx}^2+\sigma_{xy}^2}\;.
\end{equation} 
From experiment one knows that whenever 
$\rho_{xy}$ is quantised $\rho_{xx}$ gets unmeasurable small which 
in turn means that $\sigma_{xx}\to 0$ and $\rho_{xy}=1/\sigma_{yx}$. 
Therefore, to make this happen one normally concludes that the corresponding 
electronic states have to be localised. 

In the presence of frequency this argument no longer holds 
because electrons in localised states do respond to an applied
time dependent electric field giving rise to an alternating current. 
Also, both real and imaginary parts have to be considered now
\begin{equation}
\sigma_{xx}(\omega)=\sigma^R_{xx}(\omega)+\I\,\sigma^I_{xx}(\omega),
\hspace{.5cm}
\sigma_{xy}(\omega)=\sigma^R_{xy}(\omega)+\I\,\sigma^I_{xy}(\omega)\;.
\end{equation}
Assuming an isotropic system, the respective tensor components of the ac 
resistivity $\rho_{uv}(\omega)=\rho^R_{uv}(\omega)+\I\,\rho^I_{uv}(\omega)$ 
with $u,v\in\{x,y\}$ can be written as   
\begin{equation}
\rho^R_{uv}(\omega)=
\frac{\sigma^R_{vu}(\delta\sigma^2_{xx}+\delta\sigma^2_{xy})+
2\sigma^I_{vu}(\sigma^R_{xx}\,\sigma^I_{xx}+\sigma^R_{xy}\,\sigma^I_{xy})}
{(\delta\sigma^2_{xx}+\delta\sigma^2_{xy})^2
+4(\sigma^R_{xx}\,\sigma^I_{xx}+\sigma^R_{xy}\,\sigma^I_{xy})^2}
\end{equation}
\begin{equation}
\rho^I_{uv}(\omega)= 
\frac{\sigma^I_{vu}(\delta\sigma^2_{xx}+\delta\sigma^2_{xy})
+
2\sigma^R_{uv}(\sigma^R_{xx}\,\sigma^I_{xx}+\sigma^R_{xy}\,\sigma^I_{xy})}
{(\delta\sigma^2_{xx}+\delta\sigma^2_{xy})^2
+
4(\sigma^R_{xx}\,\sigma^I_{xx}+\sigma^R_{xy}\,\sigma^I_{xy})^2}
\end{equation}
with the abbreviations $\delta \sigma^2_{xx} \equiv
(\sigma^R_{xx})^2-(\sigma^I_{xx})^2$ and
$\delta \sigma^2_{xy} \equiv (\sigma^R_{xy})^2-(\sigma^I_{xy})^2$.

\section{Model and Transport theory}
We describe the dynamics of non-interacting particles moving within a
two-dimensional plane in the presence of a perpendicular magnetic
field and random electrostatic disorder potentials by a lattice model
with Hamiltonian
\begin{equation}
H=\sum_{\vec{r}}w_{\vec{r}}|\vec{r}\rangle\langle\vec{r}| - 
\hspace{-1mm}\sum_{<\vec{r}\ne \vec{r'}>}V_{\vec{r}\vec{r'}}
|\vec{r}\rangle\langle \vec{r'}|\;.
\end{equation}
The random disorder potentials associated with the lattice sites are
denoted by $w_{\vec{r}}$ with probability density distribution 
$P(w_{\vec{r}})=1/W$ within the interval $[-W/2,W/2]$, where
$W$ is the disorder strength, 
and the $|\vec{r}\rangle$ are the lattice base vectors.
The transfer terms 
\begin{equation}
V_{\vec{r}\vec{r'}}=V\exp\big(-\I\,e^2/\hbar\int_{\vec{r}}^{\vec{r'}} 
\vec{A}(\vec{l})\,\D \vec{l}\,\big)\;, 
\end{equation}
which connect only nearest neighbours on the lattice,
contain the influence of the applied magnetic field via the 
vector potential $\vec{A}(\vec{r})=(0,Bx,0)$ in their phase factors.
$V$ and the lattice constant $a$ define the units of energy and
length, respectively.

\begin{figure}[t]
\centerline{\includegraphics[width=8cm]{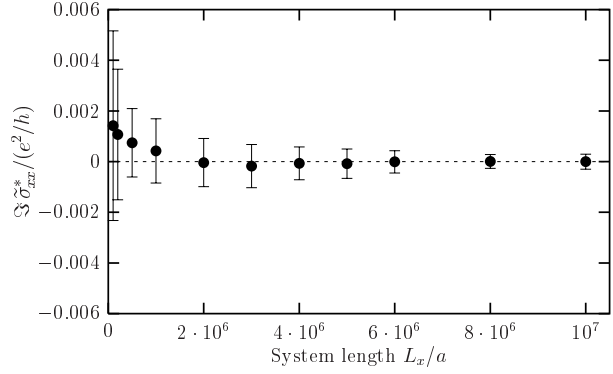}} 
\caption[]{Disorder averaged imaginary part of
$\widetilde{\sigma}^{\ast}_{xx}(E,\omega,\varepsilon)$ 
at energy $E/V=-3.35$, 
frequency $\hbar\omega/V=0.001$, and $\varepsilon/V=0.0008$
as a function of sample length $L_x=Na$. The system width is $L_y/a=32$ 
and the number of realisations amounts to 29 for $L_x/a\le 5\cdot 10^6$
and to 8 for larger lengths}  
\label{imsxxaddterm}
\end{figure}

The electrical transport is calculated within linear response theory
using the Kubo formula which allows to determine the time dependent
linear conductivity from the current matrix elements of the
unperturbed system 
\begin{eqnarray}
\sigma_{uv}(E_{\mathrm F},T,\omega)&=&\frac{\pi e^2}{\Omega}
\int_{-\infty}^{\infty}\!\D E \frac{f(E)-f(E+\hbar\omega)}{\omega} \\\nonumber
&&{}\times 
\mathrm{Tr}\big\{\hat{v}_u\delta(E_{\mathrm F}-H)
\hat{v}_v\delta(E_{\mathrm F}+\hbar\omega -H)\big\}\;,
\end{eqnarray}
where $f(E)=(\exp[(E-E_{\mathrm F})/(k_{\mathrm B}T)]-1)^{-1}$ is the Fermi 
function. The area of the system is $\Omega=L_x L_y$,
$\hat{v}_{u}=\I/\hbar [H,u]$ 
signifies the $u$-component of the velocity operator, and 
$\delta(E+\hbar\omega-H)=\I/(2\pi)(G^{\omega,+}(E)-G^{\omega,-}(E))$, 
where $G^{\omega,\pm}(E)=((E+\hbar\omega\pm\I\varepsilon)\mathrm{I}-H)^{-1}$
is the resolvent with imaginary frequency $\I\varepsilon$ and unit
matrix $\mathrm{I}$.

For finite systems at temperature $T=0$\,K, ensuring the correct order of 
limits for size $\Omega$ and imaginary frequency $\I\varepsilon$, one gets 
with $\gamma=\omega+2\I\varepsilon/\hbar$
\begin{eqnarray}
\sigma_{uv}(E_{\mathrm F},\omega) &=&
\lim_{\varepsilon\rightarrow 0}\lim_{\Omega\rightarrow\infty}
\frac{e^2}{h} \frac{1}{\Omega\hbar\omega}
\Bigg[\int_{E_F-\hbar\omega}^{E_F} \D E\,{\rm Tr}\, \Big\{(\hbar\gamma)^2 \,
[u G^{\omega,+} v G^{-}] \nonumber  \\
&&{}-
(\hbar \omega)^2 \,[u G^{\omega,+} v G^{+}]+\, 
2\I\varepsilon\,[uv(G^{\omega,+}-G^{-})]\Big\}
\nonumber \\
&&{}-
\int_{- \infty}^{E_F-\hbar \omega} \D E \, {\rm Tr }\Big\{
(\hbar \omega)^2 \,[u G^{\omega,+} v G^{+}
-u G^{\omega,-} v G^{-}]\Big\}\Bigg] \\
&=& 
\lim_{\varepsilon\rightarrow 0}\lim_{\Omega\rightarrow\infty}
\frac{e^2}{h} \frac{1}{\Omega\hbar\omega}
\Bigg[\int_{E_F-\hbar\omega}^{E_F}
\widetilde{\sigma}_{uv}(E,\omega,\varepsilon,L_x,L_y)\,\D E \nonumber\\
&&{}+
\int_{- \infty}^{E_F-\hbar \omega} 
\widetilde{\sigma}^{\ast}_{uv}(E,\omega,\varepsilon,L_x,L_y)\,\D E \Bigg]\;.
\label{cond_general}
\end{eqnarray}

\begin{figure}[t]
\centerline{\includegraphics[width=8cm]{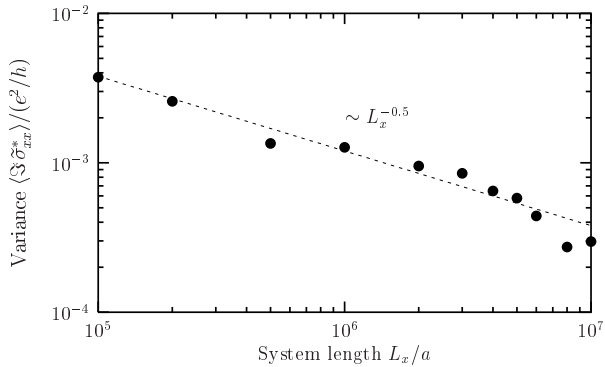}} 
\caption[]{The variance of 
$\langle\Im\,\widetilde{\sigma}^{\ast}_{xx}(E,\omega,\varepsilon)\rangle$ 
versus system length $L_x$ calculated for the averages shown in 
Fig.~\ref{imsxxaddterm} exhibiting an empirical power-law $\sim L_x^{0.5}$}
\label{varimsxxaddterm}
\end{figure}

One can show that the second integral with the limits 
($-\infty,E_{\mathrm F}-\hbar\omega$) does not contribute to the real part 
of $\sigma_{uv}(E_{\mathrm F},\omega)$ because the kernel is identical 
zero, but we were not able to proof the same also for the imaginary part. 
Therefore, using the recursive Green function method explained in the
next section, we numerically studied 
$\Im\,\widetilde{\sigma}^{\ast}_{uv}(E,\omega,\varepsilon,L_x,L_y)$ 
and found it to become very small only after disorder averaging. As an
example we show in Fig.~\ref{imsxxaddterm} the dependence of 
$\langle\Im\,\widetilde{\sigma}^{\ast}_{xx}\rangle$, averaged over up to 
29 realisations, on the length of the system for a particular energy 
$E/V=-3.35$ and width $L_y/a=32$.  
Also the variance of 
$\langle\widetilde{\sigma}^{\ast}_{xx}(E,\omega,\varepsilon)\rangle$ 
gets smaller with system length following an
empirical power law $\sim (L_x/a)^{-0.5}$ (see Fig.~\ref{varimsxxaddterm}). 
In what follows we neglect the second integral for the calculation of
the imaginary parts of the conductivities and assume that only the 
contribution of the first one with limits 
($E_{\mathrm F}-\hbar\omega,E_{\mathrm F}$) matters. 
Of course, on has to be particularly careful even if the
kernel is very small because with increasing disorder strength the
energy range that contributes to the integral (finite density of
states) tends to infinity.
Therefore, a rigorous proof for the vanishing of this integral kernel
is highly desirable.

\section{Recursive Green function method}
A very efficient method for the numerical investigation of large disordered 
chains, strips and bars that are assembled by successively adding on
slice at a time has been pioneered by MacKinnon \cite{Mac80}. 
This iterative technique relies on the property that the Hamiltonian 
$H^{(N+1)}$ of a lattice system 
containing $N+1$ slices, each a lattice constant $a$ apart, 
can be decomposed into parts that describe the system containing 
$N$ slices, $H^{(N)}$, the next slice added, $H_{N+1,N+1}$,  
and a term that connects the last slice to the rest, 
$H'_N=H_{N,N+1}+H_{N+1,N}$. 
Then the corresponding resolvent is formally equivalent to the Dyson
equation $G=G_0+G_0\widehat{V}G$ where the `unperturbed' $G_0$ represents the 
direct sum of $H^{(N)}$ and $H_{N+1,N+1}$, and $\widehat{V}$ corresponds 
to the `interaction' $H'_N$. 
The essential advantage of this method is the fact that, for a fixed width
$L_y$, the system size is increased in length adding slice by slice, 
whereas the size of the matrices to be dealt with numerically remains 
the same \cite{Mac80,MK83}.  

A number of physical quantities like localisation length 
\cite{MK81,MK82,SKM84,KPS01}, density of states \cite{SKM84,KSM84}, 
and some dc transport coefficients \cite{Mac85,SKM85,VRSM00} 
have been calculated by this technique over the years. 
Also, this method was implemented for the evaluation of the real 
part of the ac conductivity in 1d \cite{SKK83,Sas84} and 2d systems 
\cite{GB96,Gam94,GE99}.
Further efforts to included also the real and imaginary parts of the 
Hall- and the imaginary part of the longitudinal conductivity 
in quantum Hall systems were also successfully accomplished 
\cite{BS99,BS01,BS02}.

\begin{figure}[t]
\centerline{\includegraphics[width=8cm]{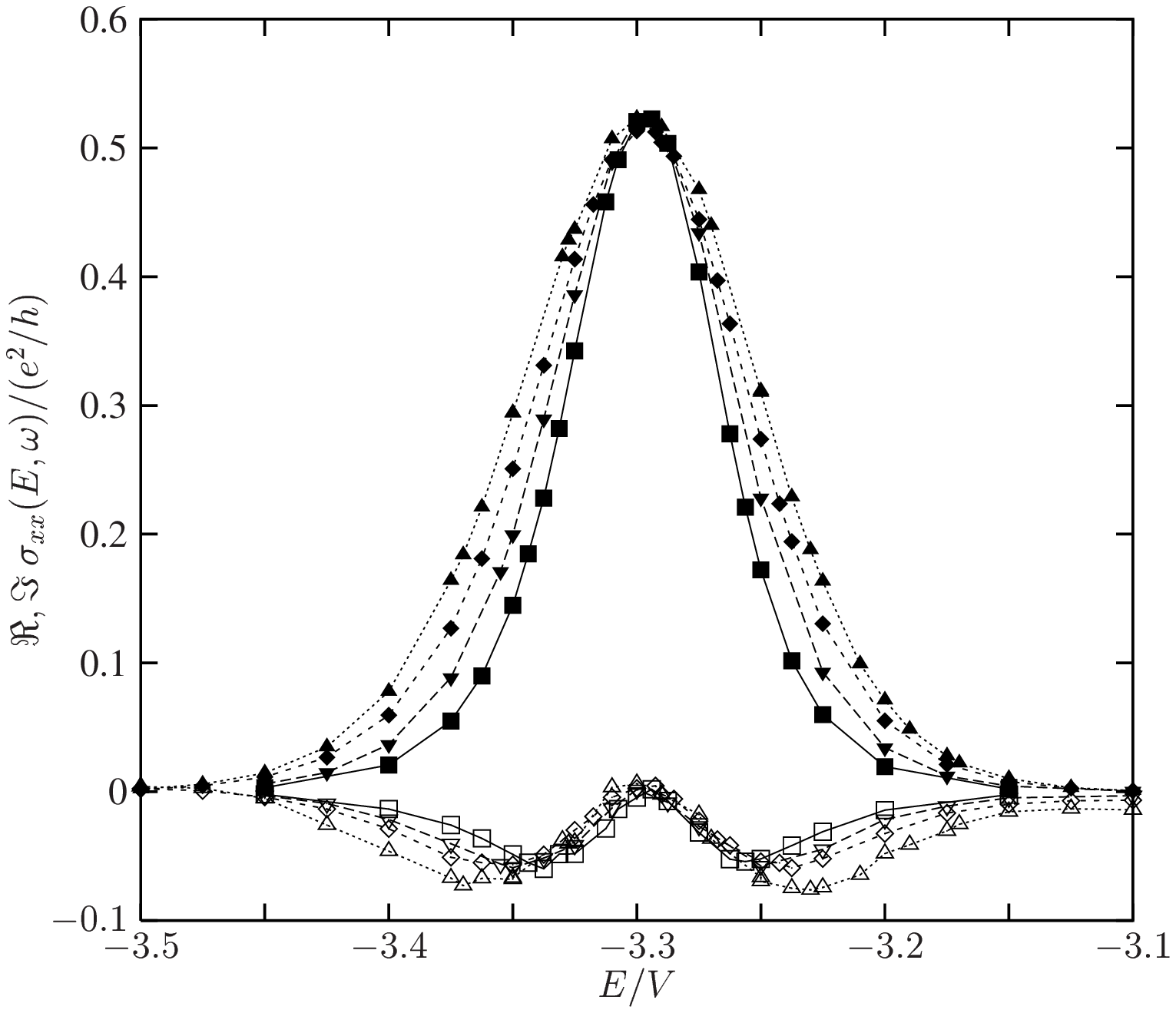}} 
\caption[]{The real and imaginary parts of
$\sigma_{xx}(E,\omega)$ as a function of energy and
frequency $\hbar\omega /V=2\cdot 10^{-4}$ (\rule{1.5mm}{1.5mm}), 
$5\cdot 10^{-4}$ (\textnormal{\DOWNarrow}), 
$1\cdot 10^{-3}$ (\rotatebox{45}{\rule{1.5mm}{1.5mm}}), 
$2\cdot 10^{-3}$ (\textnormal{\UParrow}). For comparison,
the cyclotron frequency is $\hbar\omega_c \approx 1.57\,V$ 
for $\alpha_B=1/8$ which was chosen for the magnetic flux density.
The disorder strength is $W/V=1$ and the maximal system width  
amounts to $L_y/a=96$ with periodic boundary conditions applied}
\label{RIsigxx_E_omega}
\end{figure}

The iteration equations of the resolvent matrix acting on the subspace
of such slices with indices $i,j \le N$ in the $N$-th iteration step
can be written as
\begin{eqnarray}
G_{i,j}^{\omega,\pm,(N+1)}&=&G_{i,j}^{\omega,\pm,(N)}+
G_{i,N}^{\omega,\pm,(N)}\, H_{N,N+1} \, 
G_{N+1,N+1}^{\omega,\pm,(N+1)} \, H_{N+1,N} \, 
G_{N,j}^{\omega,\pm,(N)}\nonumber\\
G_{N+1,N+1}^{\omega,\pm,(N+1)}&=&[(E+\hbar \omega\pm\I\varepsilon)
{\rm I}-H_{N+1,N+1}-H_{N+1,N} 
G_{N,N}^{\omega,\pm,(N)} H_{N,N+1}]^{-1}\nonumber \\
G_{i,N+1}^{\omega,\pm,(N+1)}&=&G_{i,N}^{\omega,\pm,(N)} \, 
H_{N,N+1} \, G_{N+1,N+1}^{\omega,\pm,(N+1)}\nonumber \\
G_{N+1,j}^{\omega,\pm,(N+1)}&=&G_{N+1,N+1}^{\omega,\pm,(N+1)} \,
H_{N+1,N}\, G_{N,j}^{\omega,\pm,(N)}\;.
\label{recur_eqn}
\end{eqnarray}

The calculation of the ac conductivities starts with the Kubo 
formula (\ref{cond_general}) by setting up a recursion 
equation for fixed energy $E$, width $L_y=Ma$, and imaginary 
frequency $\varepsilon$, which, e.g., for the longitudinal component 
reads
\begin{eqnarray}
\widetilde\sigma_{xx} (E,\omega,\varepsilon,N)&=&\frac{e^2}{h M N
a^2}S_{N}^{xx} = 
\frac{e^2}{hMNa^2}\mathrm{Tr} \bigg\{ \sum_{i,j}^{N} \Big[
(\hbar\gamma)^2 x_i G_{ij}^{\omega,+} x_j G_{ji}^{-} \nonumber \\
&&{}-(\hbar \omega)^2 x_i G_{ij}^{\omega,+} x_j G_{ji}^{+} +
2\I\varepsilon \delta_{ij} x_i^2  (G_{ij}^{\omega,+}-
G_{ji}^{-})\Big] \bigg\}\;.
\end{eqnarray}
The iteration equation for adding a new slice is given by
\begin{eqnarray}
S_{N+1}^{xx} &=& S_{N}^{xx}+ {\rm Tr} 
\big\{
A_N R_{N+1}^{-} + D_N^1 R_{N+1}^{\omega,+} D_N^2 R_{N+1}^{-}
\nonumber \\
&&{}+B_N R_{N+1}^{\omega,+} - D_N^3 R_{N+1}^{+} D_N^4 R_{N+1}^{\omega,+}  -
C_N R_{N+1}^{+} \big\}\;.
\label{newslice}
\end{eqnarray}
with $R_{N+1}^{\omega,\pm}\equiv G_{N+1,N+1}^{\omega,\pm,(N+1)}$, and
a set of auxiliary quantities as defined in the appendix. The 
coupled iteration equations and the auxiliary quantities are evaluated
numerically, the starting values are set to be zero. In addition, 
coordinate translations are required in each iteration step to keep 
the origin $x_{N+1}=0$ which then guarantees the numerical stability.   

\begin{figure}[t]
\centerline{\includegraphics[width=6.5cm]{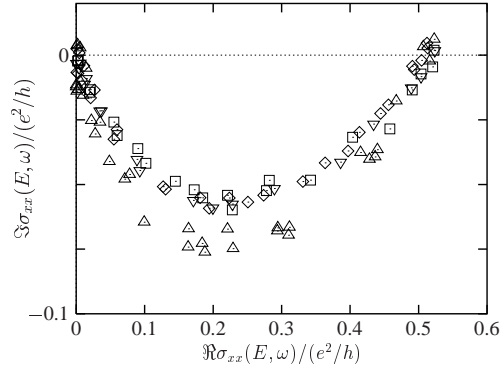}} 
\caption[]{The imaginary part of $\sigma_{xx}(E,\omega)$ 
as a function of the real part. Data are taken from 
Fig.~\ref{RIsigxx_E_omega}}
\label{imvsrealsigxx}
\end{figure}

\section{Longitudinal conductivity $\sigma_{xx}(E,\omega)$}
In this section we present our numerical results of the 
longitudinal conductivity as a function of frequency for 
various positions of the Fermi energy within the lowest Landau 
band. The real and imaginary parts of $\sigma_{xx}(E_{\mathrm F},\omega)$ 
were calculated for several frequencies, but, for the sake of legibility, 
only four of them are plotted in Fig.~\ref{RIsigxx_E_omega} versus energy.
While the real part exhibits a positive single Gaussian-like peak with
maximum $\approx 0.5\,e^2/h$ at the critical energy, the imaginary part, 
which is negative almost everywhere, has a double structure and 
vanishes near the critical point.  
$\Im\,\sigma_{xx}(E_{\mathrm F},\omega)$ almost looks like the 
modulus of the derivative of the real part with respect to energy.
Plotting $\Im\,\sigma_{xx}(E_{\mathrm F},\omega)$ as a function
of $\Re\,\sigma_{xx}(E_{\mathrm F},\omega)$ 
(see Fig.~\ref{imvsrealsigxx}) we obtain for frequencies 
$\omega < \omega^{\ast}$ a single, approximately semi-circular
curve that, up to a minus sign, closely resembles the experimental
results of Hohls \textit{et al.} \cite{HZH01}. However, for larger 
$\omega$ our data points deviate from a single curve. 

\begin{figure}[t]
\centerline{\includegraphics[width=7cm]{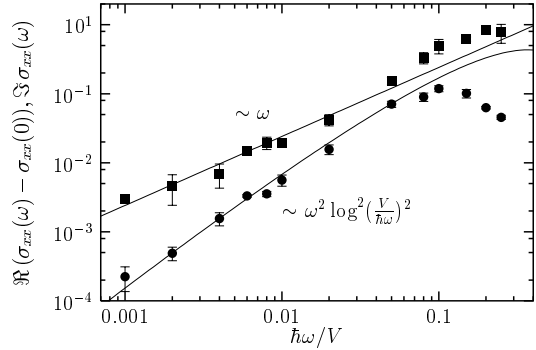}} 
\caption[]{The real ({\large $\bullet$}) and imaginary (\rule{1.5mm}{1.5mm})
part of $\sigma_{xx}(E_{\mathrm F},\omega)$ in units of $e^2/h$
at $E_{\mathrm F}/V=-3.5$ as a function of frequency. Further parameters are the
disorder strength $W/V=1.0$, system width $L_y/a=32$ and $\varepsilon/V=0.0004$}
\label{sigxxomega}
\end{figure}

\subsection{Frequency dependence of real and imaginary parts}
The behaviour of the real and imaginary part of the longitudinal ac
conductivity in the lower tail of the lowest Landau band ($E/V=-3.5$)
is shown in Fig.~\ref{sigxxomega}. We find for the imaginary part a 
linear frequency dependence for small $\omega$ which is, apart from
a minus sign, in accordance with (\ref{VEsxx}) \cite{VE90}. 
The real part can nicely be fitted to 
$\propto \omega^2\log^2[V/(\hbar\omega)]^2$ in conformance with
(\ref{VEsxx}) and $b=2$,
but disagrees with the findings in \cite{Ape89} where $b=1$ was
proposed.

\subsection{Behaviour of the maximum of $\sigma_{xx}(\omega)$}
\begin{figure}[t]
\centerline{\includegraphics[width=7cm]{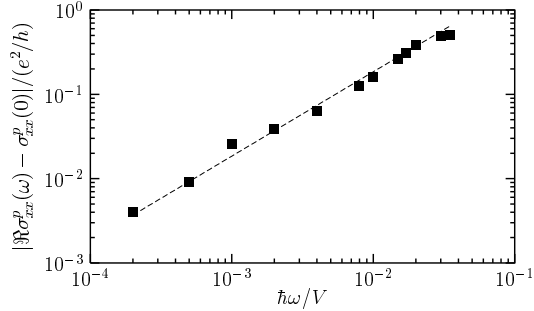}} 
\caption[]{The non-Drude decrease of the peak value of the
longitudinal conductivity as a function of frequency.
A linear behaviour of 
$|\Re\,\sigma^{\mathrm p}_{xx}(\omega)-\sigma^{\mathrm p}_{xx}(0)|$ 
is clearly observed with $\sigma^{\mathrm p}_{xx}(0)/(e^2/h)=0.512$
using the following parameters $E/V=-3.29$, $W/V=0.1$, $L_y/a=32$, 
$\varepsilon/V=0.0004$}
\label{xxWp1E3p29B32O}
\end{figure}

The frequency dependence of the real part of the longitudinal
conductivity peak value was already investigated in \cite{GB96}
where for long-range correlated disorder potentials a non-Drude-like
decay was observed. We obtained a similar behaviour also 
for spatially uncorrelated disorder potentials in a lattice model
\cite{BS01}. In Fig.~\ref{xxWp1E3p29B32O} the difference 
$|\Re\,\sigma^{\mathrm p}_{xx}(\omega)-\sigma^{\mathrm p}_{xx}(0)|$
is plotted versus frequency in a double logarithmic plot from which
a linear relation can be discerned.
A linear increase with frequency was found for the imaginary part 
of the longitudinal conductivity at the critical point as well 
\cite{BS01}.
The standard explanation for the non-Drude behaviour in terms of 
long time tails in the velocity correlations, which were shown to
exist in a QHE system \cite{GB96,BGK97}, seems not to be adequate 
in our case. For the uncorrelated disorder potentials considered 
here, it is not clear whether the picture of electron motion 
along equipotential lines, a basic ingredient for the arguments in 
\cite{GB96}, is appropriate.  

\begin{figure}[b]
\centerline{\includegraphics[width=7cm]{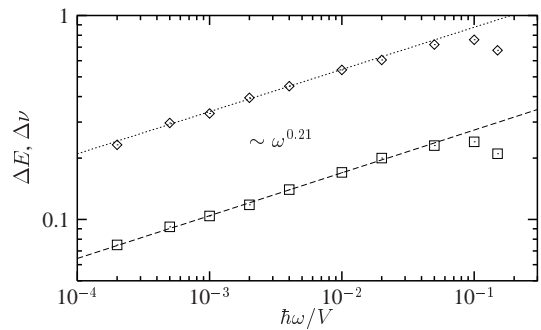}} 
\caption[]{Frequency scaling of the $\sigma_{xx}(\omega)$ peak width. 
The width in energy $\Delta E$ ($\Box$) and the width in filling
factor $\Delta \nu$ ({\large$\diamond$}) show a power-law 
$\Delta \sim \omega^{\kappa}$ with an exponent 
$\kappa=(\mu z)^{-1}=0.21$}
\label{sigxx_fs_E_nu}
\end{figure}

\subsection{Scaling of the $\sigma_{xx}(\omega)$ peak width}
The scaling of the width of the conductivity peaks with frequency is
shown in Fig.~\ref{sigxx_fs_E_nu} where both the $\sigma_{xx}(\omega)$
peak width expressed in energy and, due to the knowledge of the density 
of states, in filling factor are shown to follow a power-law 
$\Delta \sim \omega^{\kappa}$ with $\kappa=(\mu z)^{-1}=0.21$ \cite{BS99}.
Taking $\mu=2.35$ from \cite{Huc92} we get $z=2.026$ close to what 
is expected for non-interacting electrons. 
Therefore, the result reported in \cite{AL96} 
seems to be doubtful. However, spatial correlations in the disorder
potentials as considered in \cite{AL96} may influence the outcome.
Alternatively, one could fix $z=2$ and obtain $\mu=2.38$ in close
agreement with the results from numerical calculations of the scaling 
of the static conductivity \cite{GB94} or the localisation length 
\cite{Huc92}. 

The experimentally observed values $\kappa\simeq 0.42$ \cite{ESKT93} and
$\kappa=0.5\pm 0.1$ \cite{HZHMDK02} are larger by a factor of about 2.
This is usually attributed to the influence of electron-electron 
interactions ($z=1$) which were neglected in the numerical investigations. 

\begin{figure}[t]
\centerline{\includegraphics[width=5.77cm]{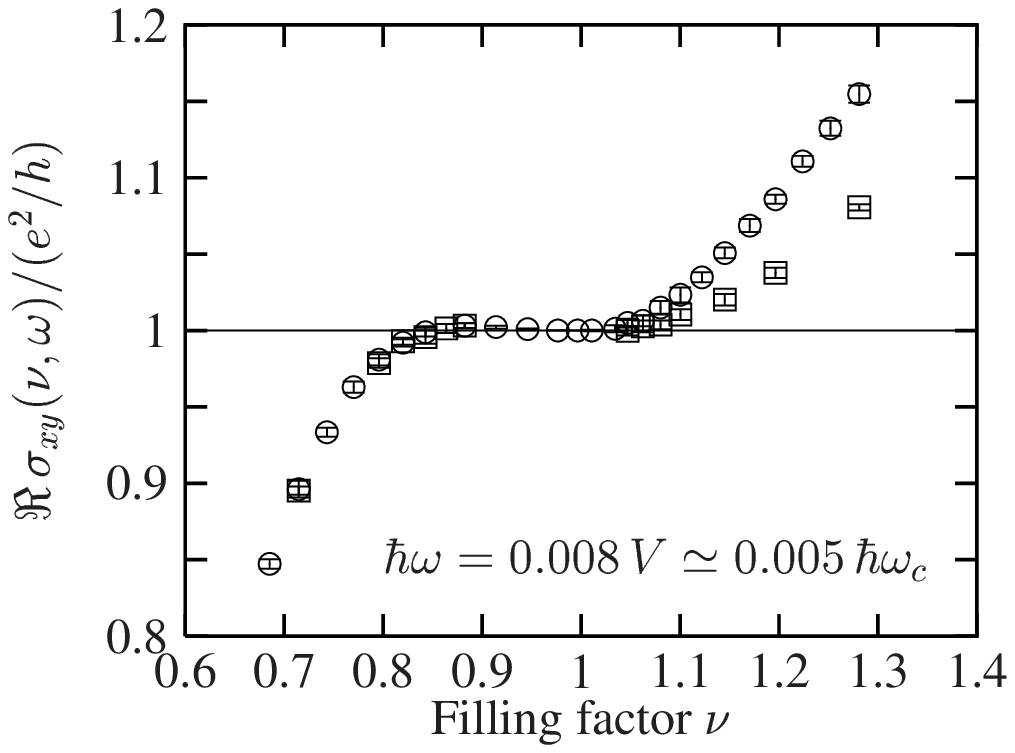}%
\hspace{.35cm}\includegraphics[width=5.5cm]{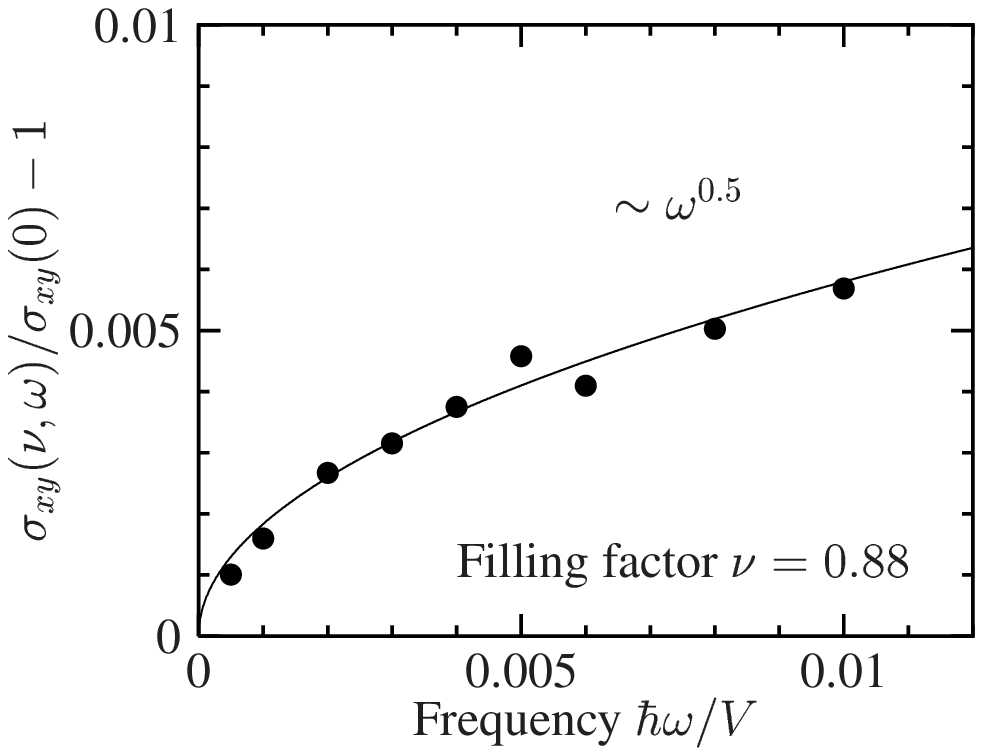}} 
\caption[]{Frequency dependent Hall-conductivity: 
The figure on the l.h.s shows $\sigma_{xy}(\omega)$
versus filling factor for two system sizes $L_y/a=64$ ({\large$\circ$}) 
and $L_y/a=128$ ($\Box$). On the r.h.s., the relative 
deviation of the Hall-conductivity from the dc value, 
$\sigma_{xy}(\omega)/\sigma_{xy}(0)-1$, 
is plotted as a function of frequency for $\nu=0.88$}
\label{sigxy_pics}
\end{figure}

\section{Frequency dependent Hall-conductivity}
The Hall-conductivity due to an external time dependent electric field 
as a function of filling factor $\nu$ is shown in Fig.~\ref{sigxy_pics} 
for system widths $L_y/a=64$ and $L_y/a=128$, respectively. While 
$\sigma_{xy}(\omega)$ has already converged for $E_{\mathrm F}$ lying
in the upper tail of the lowest Landau band a pronounced shift can be 
seen in the lower tail of the next Landau band. For $\nu = 1.3$ a system 
width of at least $L_y/a=192$ was necessary for $\sigma_{xy}(\omega)$ 
to converge. This behaviour originates in the exponential increase of 
the localisation length with increasing Landau band index. Due to the
applied frequency $\hbar\omega/V=0.008$ the $\sigma_{xy}$ plateau is 
not flat, but rather has a parabola shape near the minimum at $\nu = 1.0$,
similar to what has been observed in experiment \cite{SMCHAP02}. 
An example of the deviation of $\sigma_{xy}(\omega)$ from its quantised 
dc value is shown on the right hand side of Fig.~\ref{sigxy_pics} where 
$\sigma_{xy}(\omega)/\sigma_{xy}(0)-1$ is plotted versus frequency
for filling factor $\nu=0.88$.
A power-law curve $\sim \omega^{0.5}$ can be fitted to the data points. 
Using this empirical relation, we find a relative deviation of the
order of $5\cdot 10^{-6}$ when extrapolated down to to 1\,kHz, 
the frequency usually applied in metrological experiments 
\cite{Del93,Del94,CHK99,SMCHAP02}. Therefore, there is no quantisation
in the neighbourhood of integer filling even in an ideal 2d electron
gas without contacts, external leads, and other experimental imperfections.
Recent calculations, however, show that
this deviation can be considerably reduced, even below $1\cdot 10^{-8}$, 
if spatially correlated disorder potentials are considered in the
model \cite{Sch03}.

\section{Conclusions}
The frequency dependent electrical transport in integer quantum Hall 
systems
has been reviewed and the various theoretical developments have been 
presented. Starting from a linear response expression a method has 
been demonstrated which is well suited for the numerical evaluation of 
the real and imaginary parts of both the time dependent longitudinal 
and the Hall-conductivity. In contrast to the analytical approaches, 
no further approximations or restrictions such as the position of the 
Fermi energy have to be considered. 

We discussed recent numerical results in some detail with particular 
emphasis placed on the frequency scaling of the peak width of the 
longitudinal conductivity emerging at the quantum critical points,
and on the quantisation of the ac Hall-conductivity at the plateau.
As expected, the latter was found to depend on the applied frequency. 
The extrapolation of our calculations down to low frequencies resulted
in a relative deviation
of $5\cdot 10^{-6}$ at $\sim 1\,$kHz when spatially uncorrelated
disorder potentials are considered. Disorder potentials with spatial 
correlations, likely to exist in real samples, will probably reduce 
this pronounced frequency effect.

Our result for the frequency dependence of the $\sigma_{xx}$ peak width 
showed power-law scaling, $\Delta \sim \omega^{\kappa}$, where
$\kappa=(\mu z)^{-1}=0.21$ as expected for non-interacting electrons.   
Therefore, electron-electron interactions have presumably to be taken 
into account to explain the experimentally observed $\kappa\approx 0.5$
Also, the influence of the spatial correlation of the disorder
potentials may influence the value of $\kappa$.

The frequency dependences of the real and imaginary parts of the longitudinal 
conductivities, previously obtained analytically for Fermi energies lying 
deep down in the lowest tail of the lowest Landau band, have been confirmed 
by our numerical investigation. However, the quadratic behaviour found
at low frequencies for the real part of $\sigma_{xx}(\omega)$ has to
be contrasted with the linear frequency dependence that has been proposed for 
hopping conduction.

Finally, a non-Drude decay of the $\sigma_{xx}$ peak value with frequency,
$\sigma^{p}_{xx}(0)-\sigma^{\mathrm p}_{xx}(\omega) \propto \omega$,
as reported earlier for correlated disorder potentials, has been observed
also in the presence of uncorrelated disorder potentials. A convincing
explanation for the latter behaviour is still missing.

\newpage
\section*{Appendix}
The iteration equations of the auxiliary quantities (required in (\ref{newslice})) 
can be written as \cite{BS02} with $V_N\equiv H_{N,N+1}=H^{\dagger}_{N+1,N}$
\begin{eqnarray}
A_{N+1} &=& V_{N+1}^{\dag} R_{N+1}^{-} \big[
A_N + D_N^1 R_{N+1}^{\omega,+} D_N^2 \big] R_{N+1}^{-} V_{N+1}\\
B_{N+1} &=& V_{N+1}^{\dag} R_{N+1}^{\omega,+}\big[ 
B_N - D_N^3 R_{N+1}^{+} D_N^4 + 
D_N^2 R_{N+1}^{-} D_N^1 
\big] R_{N+1}^{\omega,+} V_{N+1}\\
C_{N+1} &=& V_{N+1}^{\dag} R_{N+1}^{+} \big[ C_N +
D_N^4 R_{N+1}^{\omega,+} D_N^3 \big] R_{N+1}^{+} V_{N+1}\\
F_{N+1} &=& V_{N+1}^{\dag} R_{N+1}^{\omega,+} \big[ F_N +
D_N^3 R_{N+1}^{+} D_N^4 \big] R_{N+1}^{\omega,+} V_{N+1}\\
G_{N+1} &=& V_{N+1}^{\dag} R_{N+1}^{-} \big[ G_N +
D_N^{10} R_{N+1}^{\omega,-} D_N^{11} \big] R_{N+1}^{-} V_{N+1}\\
H_{N+1} &=& V_{N+1}^{\dag} R_{N+1}^{\omega,-} \big[ H_N +
D_N^{11} R_{N+1}^{-} D_N^{10} \big] R_{N+1}^{\omega,-} V_{N+1}\\
D_{N+1}^1 &=& V_{N+1}^{\dag} R_{N+1}^{-} 
D_N^1
R_{N+1}^{\omega,+}V_{N+1}\\
D_{N+1}^2 &=& V_{N+1}^{\dag} R_{N+1}^{\omega,+} 
D_N^2
R_{N+1}^{-} V_{N+1}\\
D_{N+1}^3 &=& V_{N+1}^{\dag} R_{N+1}^{\omega,+} 
D_N^3
R_{N+1}^{+} V_{N+1}\\
D_{N+1}^4 &=& V_{N+1}^{\dag} R_{N+1}^{+} 
D_N^4 
R_{N+1}^{\omega,+} V_{N+1}\\
D_{N+1}^5 &=& V_{N+1}^{\dag} R_{N+1}^{-} 
D_N^5
R_{N+1}^{-} V_{N+1}\\
D_{N+1}^6 &=& V_{N+1}^{\dag} R_{N+1}^{+} 
D_N^6 
R_{N+1}^{+} V_{N+1}\\
D_{N+1}^{10} &=& V_{N+1}^{\dag} R_{N+1}^{-} 
D_N^{10} 
R_{N+1}^{\omega,-} V_{N+1}\\
D_{N+1}^{11} &=& V_{N+1}^{\dag} R_{N+1}^{\omega,-} 
D_N^{11} 
R_{N+1}^{-} V_{N+1}\\
D_{N+1}^{14} &=& V_{N+1}^{\dag} R_{N+1}^{\omega,-} 
D_N^{14} 
R_{N+1}^{\omega,-} V_{N+1}\\
D_{N+1}^{15} &=& V_{N+1}^{\dag} R_{N+1}^{\omega,+} 
D_N^{15} 
R_{N+1}^{\omega,+} V_{N+1}\\
D_{N+1}^{16} &=& V_{N+1}^{\dag} R_{N+1}^{-} 
D_N^{16} 
R_{N+1}^{-} V_{N+1}\\
E_{N+1}^1 &=& V_{N+1}^{\dag} R_{N+1}^{+} 
\Big(
E_N^1 + (\hbar \omega) I
\Big)
R_{N+1}^{+} V_{N+1}\\
E_{N+1}^2 &=& V_{N+1}^{\dag} R_{N+1}^{-} 
\Big(
E_N^2 + (\hbar \omega) I
\Big)
R_{N+1}^{-} V_{N+1}\\
E_{N+1}^3 &=& V_{N+1}^{\dag} R_{N+1}^{\omega,+} 
\Big(
E_N^3 + (\hbar \omega) I
\Big)
R_{N+1}^{\omega,+} V_{N+1}\\
E_{N+1}^4 &=& V_{N+1}^{\dag} R_{N+1}^{\omega,-} 
\Big(
E_N^4 + (\hbar \omega) I
\Big)
R_{N+1}^{\omega,-} V_{N+1}\;,
\end{eqnarray}
where the auxiliary quantities are defined by
\begin{eqnarray}
A_N &=& V_{N}^{\dag} \bigg( \sum_{i,j}^{N} G_{Ni}^{-}x_i
\big[ (\hbar\gamma)^2 G_{ij}^{\omega,+} - 2\I\varepsilon \delta_{ij} {\rm I} \big]
x_j G_{jN}^{-}\bigg)V_{N}\\
B_N &=&V_{N}^{\dag} \bigg( \sum_{i,j}^{N}G_{Ni}^{\omega,+} x_i
\big[ (\hbar\gamma)^2 G_{ij}^{-} + 2\I\varepsilon \delta_{ij} {\rm I}+
(\hbar \omega)^2 G_{ij}^{+}\big] x_j G_{jN}^{\omega,+} \bigg)V_{N}\\
C_N &=& V_{N}^{\dag} \bigg( \sum_{i,j}^{N}
(\hbar \omega)^2 G_{Ni}^{+} x_i G_{ij}^{\omega,+} x_j 
G_{jN}^{+}\bigg) V_{N}\\
F_N &=& V_{N}^{\dag} \bigg( \sum_{i,j}^{N}
(\hbar \omega)^2 G_{Ni}^{\omega,+} x_i G_{ij}^{+} x_j 
G_{jN}^{\omega,+}\bigg) V_{N}\\
G_N &=& V_{N}^{\dag} \bigg( \sum_{i,j}^{N}
(\hbar \omega)^2 G_{Ni}^{-} x_i G_{ij}^{\omega,-} x_j 
G_{jN}^{-}\bigg) V_{N}\\
H_N &=& V_{N}^{\dag} \bigg( \sum_{i,j}^{N}
(\hbar \omega)^2 G_{Ni}^{\omega,-} x_i G_{ij}^{-} x_j 
G_{jN}^{\omega,-}\bigg) V_{N}\\
D_N^1 &=& V_{N}^{\dag} \bigg( \sum_{i}^{N}
(\hbar\gamma) G_{Ni}^{-} x_i G_{iN}^{\omega,+}\bigg)V_{N}\\
D_N^2 &=& V_{N}^{\dag} \bigg( \sum_{i}^{N}
(\hbar\gamma) G_{Ni}^{\omega,+} x_i G_{iN}^{-}\bigg)V_{N}\\
D_N^3 &=& V_{N}^{\dag} \bigg( \sum_{i}^{N}
(\hbar \omega) G_{Ni}^{\omega,+} x_i G_{iN}^{+}\bigg)V_{N}\\
D_N^4 &=& V_{N}^{\dag} \bigg( \sum_{i}^{N}
(\hbar \omega) G_{Ni}^{+} x_i G_{iN}^{\omega,+}\bigg)V_{N}\\
D_N^5 &=& V_{N}^{\dag} \bigg( \sum_{i}^{N}
(\hbar\omega ) G_{Ni}^{-} x_i  G_{iN}^{-}\bigg)V_{N}\\
D_N^6 &=& V_{N}^{\dag} \bigg( \sum_{i}^{N}
(\hbar \omega) G_{Ni}^{+} x_i  G_{iN}^{+}\bigg)V_{N} \\
D_N^{10} &=& V_{N}^{\dag} \bigg( \sum_{i}^{N}
(\hbar \omega) G_{Ni}^{-} x_i  G_{iN}^{\omega,-}\bigg)V_{N} \\
D_N^{11} &=& V_{N}^{\dag} \bigg( \sum_{i}^{N}
(\hbar \omega) G_{Ni}^{\omega,-} x_i  G_{iN}^{-}\bigg)V_{N} \\
D_N^{14} &=& V_{N}^{\dag} \bigg( \sum_{i}^{N}
(\hbar \omega) G_{Ni}^{\omega,-} x_i  G_{iN}^{\omega,-}\bigg)V_{N} \\
D_N^{15} &=& V_{N}^{\dag} \bigg( \sum_{i}^{N}
(\hbar \omega) G_{Ni}^{\omega,+} x_i  G_{iN}^{\omega,+}\bigg)V_{N} \\
D_N^{16} &=& V_{N}^{\dag} \bigg( \sum_{i}^{N}
(\hbar \omega) G_{Ni}^{-} x_i  G_{iN}^{-}\bigg)V_{N} \\
E_N^1 &=& V_{N}^{\dag} \bigg( \sum_{i}^{N}
(\hbar \omega) G_{Ni}^{+} G_{iN}^{+}\bigg)V_{N} \\
E_N^2 &=& V_{N}^{\dag} \bigg( \sum_{i}^{N}
(\hbar\omega ) G_{Ni}^{-} G_{iN}^{-}\bigg)V_{N}\\
E_N^3 &=& V_{N}^{\dag} \bigg( \sum_{i}^{N}
(\hbar \omega) G_{Ni}^{\omega,+} G_{iN}^{\omega,+}\bigg)V_{N} \\
E_N^4 &=& V_{N}^{\dag} \bigg( \sum_{i}^{N}
(\hbar \omega) G_{Ni}^{\omega,-} G_{iN}^{\omega,-}\bigg)V_{N}\;.
\end{eqnarray}
For the translation of the coordinates $x_i \to x_i - a$ one gets
\begin{eqnarray}
A_N^{'} &=& A_N + H_N^1 - 2 a D_N^5 + a^2 E_N^2 
+ a H_N^3\\
B_N^{'} &=& B_N + H_N^2 - H_N^1 + a (H_N^4-H_N^3) \\
C_N^{'} &=& C_N + H_N^2 - 2 a D_N^6 + a^2 E_N^1
+ a H_N^4 \\
F_N^{'} &=& F_N - a D_N^4 + 2 a D_N^{15}
- a D_N^3 - a H_N^4 - a^2 E_N^3 \\
G_N^{'} &=& G_N + a (D_N^{10} + D_N^{11}) - 2 a D_N^{16} + a^2 E_N^2
+ a H_N^5 \\
H_N^{'} &=& H_N -a D_N^{10} + 2 a D_N^{14}
- a D_N^{11} - a H_N^5 - a^2 E_N^4 \\
D_N^{1/2'} &=& D_N^{1/2} + H_N^3 \\
D_N^{3/4'} &=& D_N^{3/4} + H_N^4 \\
D_N^{5'} &=&D_N^{5} - a E_N^2\\ 
D_N^{6'} &=&D_N^{6} - a E_N^1 \\
D_N^{10/11'} &=&D_N^{10/11} + H_N^5 \\
D_N^{14'} &=&D_N^{14} - a E_N^4 \\
D_N^{15'} &=&D_N^{15} - a E_N^3 \\
D_N^{16'} &=&D_N^{16} - a E_N^2\;,
\end{eqnarray}
with the following abbreviations
\begin{eqnarray}
H_N^1 &=& a (D_N^{1}+D_N^{2})\\
H_N^2 &=& a (D_N^{3}+D_N^{4})\\
H_N^3 &=& a V_{N}^{\dag} (R_{N}^{\omega,+} -R_{N}^{-})V_{N} \\
H_N^4 &=& a V_{N}^{\dag} (R_{N}^{\omega,+} -R_{N}^{+})V_{N} \\
H_N^5 &=& a V_{N}^{\dag} (R_{N}^{\omega,-} -R_{N}^{-})V_{N}\;.
\end{eqnarray} 

\newpage


%

\end{document}